# Internal advection dynamics in sessile droplets depend on the curvature of superhydrophobic surfaces

**Gargi Khurana[1], A R Harikrishnan[2], Vivek Jaiswal[1] and Purbarun Dhar[1, *]**

[1] Department of Mechanical Engineering, Indian Institute of Technology Ropar,

Rupnagar–140001, India

[2] Department of Mechanical Engineering, Indian Institute of Technology Madras,

Chennai–600036, India

*Corresponding author*

E–mail: purbarun@iitrpr.ac.in

Phone: +91–1881–24–2173

**Abstract**

The article shows experimentally that the internal circulation velocity and patterns in sessile droplets on superhydrophobic surfaces is a function of the surface curvature. Particle Image Velocimetry (PIV) analysis reveals that increasing convexity deteriorates the advection velocity whereas concavity augments the same. A scaling model analysis based on the effective curvature modulated change in wettability is found to predict the phenomenon, but weakly. Potential flow theory is appealed to and the curvatures are approximated as wedges with the rested droplet engulfing them in part. The spatially averaged experimental velocities are found to conform to the mathematical predictions. The present study may have strong implications in thermofluidics transport phenomena at the microscale.

*Keywords:* Superhydrophobicity, droplets, potential flow, internal circulation, PIV

*Introduction* – Fluid dynamics, thermal and species transport in droplets has been an area of tremendous academic interest, since droplet dynamics is essential in several scientific and engineering applications. Examples involve combustion engines[1, 2], microfluidic biological manipulation[3, 4]; spray coating[5] and drug administering[6], printing technology[7, 8], agriculture[9, 10], cooling technologies[11], and so forth. Droplets are broadly classified into the pendent (suspended) and sessile (seated). The sessile droplets pose a more complex condition as its shape and size is determined by the equilibrium of three phases (liquid, surrounding gas and solid surface). Consequently, thermofluidics in sessile droplets, typically droplet impact and spreading events[12-14], thermal transport[15] and mass transfer[16], has received extensive attention among academicians. Additionally, the sessile droplet dynamics are strongly governed by the wettability of the surface, and research in the direction ranging from superhydrophobic (SH) [17, 18] to superhydrophilicity[19] have received immense focus in the last few years.

One of the more interesting phenomena in sessile droplets, typically those resting on SH surfaces, is the presence of prominent internal fluid circulation[20-22]. This is brought about by the Marangoni stresses generated across the droplet interface due to proximity to surface molecules which have aversion towards water molecules. The stresses generated, along with mass continuity principle, leads to advective motion within an otherwise stationary droplet. Interestingly, pendent droplets also exhibit internal circulation when thermal or solutal Marangoni advection is forcefully imposed[23, 24]. The motivation of the present study derives clue from reports that curvature of a surface essentially modulates the wettability of a droplet[25, 26]. The modulated wettability would thus be expected to govern the strength and pattern of internal circulation in sessile droplets seated on curved SH surfaces. The present article describes a novel phenomenon corroborating this theory and experimentally and analytically illustrates the curvature dependent modulation of internal circulation dynamics on SH surfaces.

*Experimental methods* - The experimental setup has been illustrated in Figure 1. It consists of a digitized droplet dispensing mechanism (Holmarc Optomechatronics Ltd. India) connected to a glass chromatography syringe (Unitech, India) capable of dispensing accurate to ± 0.1 μl. Sessile droplets of average volume ~ 15 μl are dispensed onto the SH surfaces. The surfaces used are of three different types, viz. cylindrical rods (stainless steel) (promotes single positive curvature), spheres (stainless steel) (dual positive curvature), and concave semi-cylindrical grooves (stainless steel, manufactured employing wire-cut micro electro discharge machining (EDM)) (single negative curvature). The control case is a flat (stainless steel) SH surface (zero curvature). All surfaces were rendered SH by spray coating with a commercial SH twin component infusion (Rust-oleum Never Wet, USA). Since deionized water exhibits weak internal circulation on SH surfaces[24], 0.05 M aqueous NaCl solution (analytical grade, Sigma Aldrich, India) is employed as it enhances internal circulation due to solutal Marangoni advection without appreciable increase in surface tension[24]. The static contact angle of the solution on flat SH surface is determined to be ~ 157 ± 1.5° and roll-off angle as < 3°. The experimental setup was housed in an acrylic chamber to eliminate convective disturbances. A digital thermometer and hygrometer was employed to monitor conditions within the chamber. All experiments were performed at 30 ± 1 °C and relative humidity of 48 ± 2.5 %.

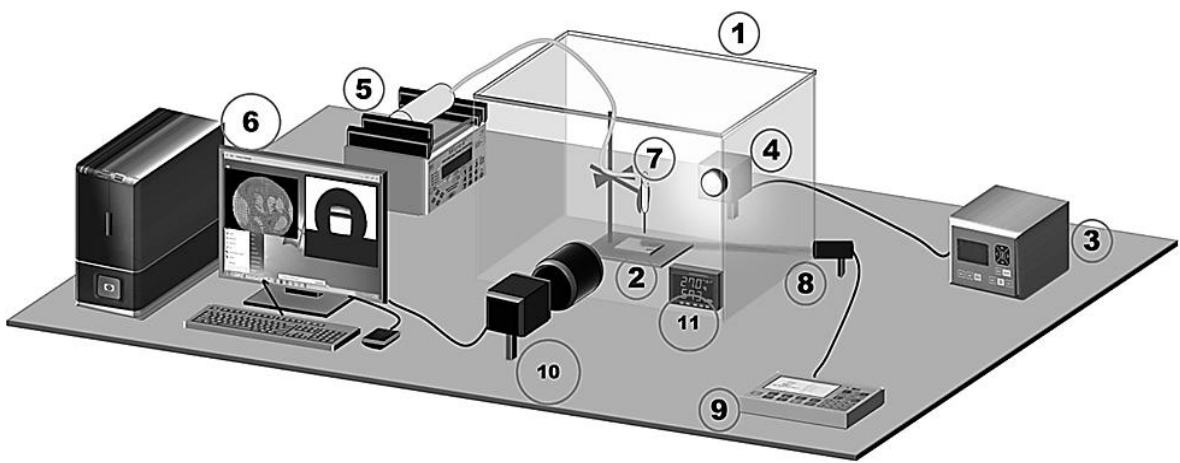

**Figure 1:** Schematic of the experimental setup, 1) acrylic chamber 2) SH surface 3) LED array controller 4) LED backlight 5) droplet dispenser 6) data acquisition computer 7) syringe 8) laser 9) laser controller 10) CCD camera with lens assembly and 11) digital thermometer and hygrometer.

The salt solution is seeded with neutrally buoyant (with water at 300 K), fluorescent polystyrene particles (20 μm diameter, Cospheric LLC, USA) for flow visualization and quantification using Particle Image Velocimetry (PIV). A CCD camera (1280 x 960 pixels resolution at 20 fps, Holmarc Optomechatronics Ltd. India) with a microscopic lens assembly and an intensity controllable LED backlight (DPLED 5S, China) is used to focus the droplet image using an image acquisition software. A continuous wave laser of wavelength 532nm and 10 mW peak power (Roithner GmbH, Germany) is used as the source of illumination for the PIV. A laser sheet of ~ 0.5 mm thickness is used to illuminate the droplet interior using sheet optics. Once the camera is focussed onto the droplet, the LED backlight is extinguished and the laser is used as the light source. A camera resolution of ~ 120pixels/mm was employed for the PIV and images were recorded at 10 fps. The PIV data acquisition was performed within 3-5 minutes of placing the droplet, thereby exterminating possibilities of evaporation driven internal advection in addition to the surface energy driven advection[24, 23]. For post-processing, a cross-correlation, four-pass algorithm with subsequent interrogation windows of 64, 32, 16 and 8 pixels was used to obtain high signal to noise ratio in the velocity quantification. The data was processed for typically 500 images for each case in the open source code PIVlab and the time-averaged, mean spatial velocity contours and vectors were obtained.

*Results and analysis* – Convex cylinders of radii 5, 3.5 and 2 mm, convex spheres of radii 4.75, 3.5 and 2.35 mm, and concave cylindrical grooves of radii 3, 2 and 1 mm were studied. The droplet was carefully seated axisymmetric on the curved surfaces and flow visualization was performed. Fig. 2 illustrates the time average, spatial velocity contours and vector distribution for variant convex curved systems compared to the flat SH case. It is observable that while the circulation pattern is more or less retained throughout the droplet, the overall strength of the mean circulation velocity, as well as the maximum velocity within the droplet diminishes as a function of increasing curvature. Additionally, it is observed that the reduction in velocity is further augmented in case of spheres than the cylinders. This can be attributed to the fact that the two curvatures of spheres promote greater resistance to the flow than the single curvature of cylinders. For concave cylinders (fig. 3), the opposite is observed and augmentation in velocity is noted. The velocity increases with increased concave curvature. It is noteworthy however, that for droplet radius equal to concave groove radius (fig. 3 (c)) the velocities drastically reduce. This is attributed to the frictional dissipation at the wall caused due to contact of droplet outer surface to the groove inner walls.

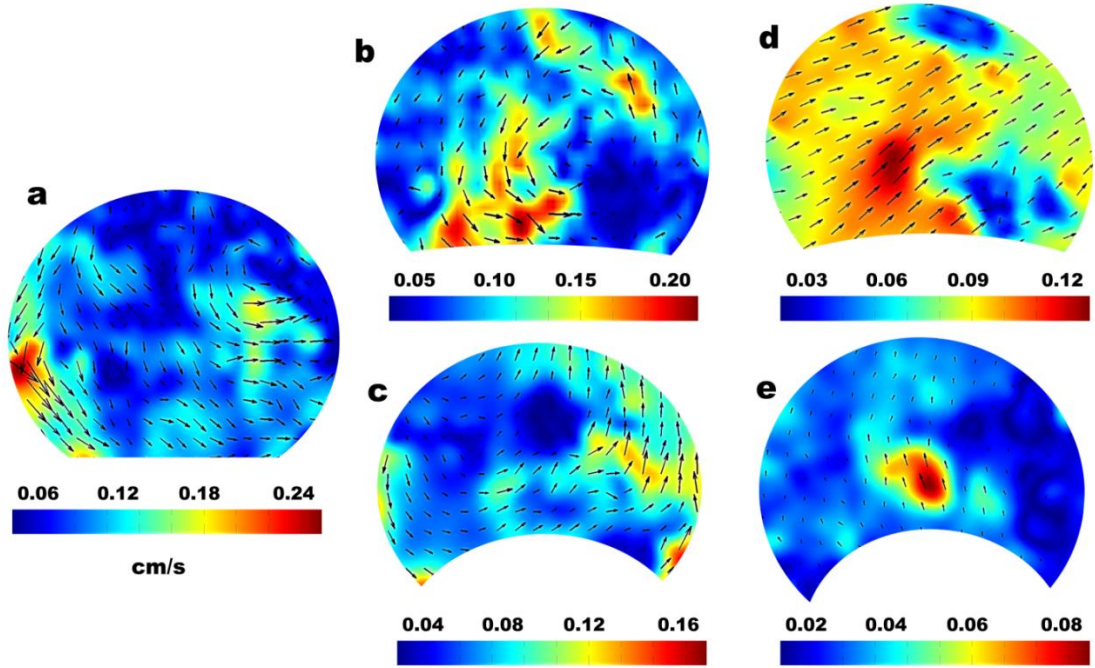

**Figure 2:** Velocity contours and vector plots for internal circulation in droplets on (a) flat (b) 5 mm convex cylinder (c) 2 mm convex cylinder (d) 4.75 mm concave cylinder and (e) 2.35 mm concave cylinder SH surfaces.

The mechanism behind the change in the average velocity of internal circulation needs further discussion. On a SH surface, the water molecules neighboring the surface molecules are continuously repelled off, and neighboring molecules move in to satisfy mass continuity within the droplet. This typical movement leads to a permanent drift or circulation within the droplet on SH surfaces; a phenomenon absent in hydrophilic surfaces due to affinity of the water molecules for the surface molecules. In case of curved interfaces, the effective contact angle at the three phase contact line changes, thereby modifying the effective wettability of the surface. The effective contact angle on such curvatures can be deduced from image analysis and geometrical considerations, along the lines detailed in Fig. S3[a]. The effective contact angle for a droplet on a curved superhydrophobic (SH) surface is different from its static contact angle on a flat SH surface. This effective change in contact angle essentially signifies that the wettability on a curved surface is a function of the curvature[25]. As per fig S3 [a], let the static contact angle on a SH surface be θ. Let a droplet of same volume be now rested onto a surface with a known curvature (inverse of radius of the curve, a). The convex curvature is traditionally treated as positive, whereas the concave curvature as negative and the sign needs to be incorporated in the curve radius (a) [a]. In the event the frontal projected length of the contact diameter (c)[a] can be determined, the effective contact angle (θ')[a] can be mathematically expressed as

$$\theta' = \theta - sin^{-1}(c/a) \qquad (1)$$

It may be readily observed that with increasing positive curvature (or convexity) or decreasing curve radius, the difference between the effective contact angle and true contact angle increases, indicating higher effective wetting. From an intuitive standpoint also this mechanism is correct. A droplet sitting on a spherical cap engulfs a larger contact surface compared to its flat counterpart, thereby making if a more wetted contact. On the contrary, increased concavity reduces the contact area of the droplet, leading to augmented effective contact angle and improved hydrophobicity. The deteriorated hydrophobicity in convex SH surface signifies that it has shifted a little closer to the hydrophilic regime than the flat case, which could explain the suppression in circulation velocity (droplets exhibit stagnant interiors on hydrophilic surfaces) and the same logic explains augmented circulation velocity with increasing concavity. Thereby, it can be proposed that the ratio of contact angles with respect to the flat case could be a measure of the change in wettability (the ratio, $\theta'/\theta$ will be $> 1$ for increasing concavity and $< 1$ for increasing convexity). Based on order of magnitude scaling, it could be argued that the ratio of average circulation velocity in curved case ($U_c$) to that of the flat case ($U_f$) will scale as $(U_c/U_f) \sim (\theta'/\theta)^n$, where $n$ is a real number. However, analysis shows that the typical scaling does not hold true and the experimental observations do not comply with the scaling over the whole curvature spectrum. Thereby, while the effective wettability ratio might be important, is alone cannot predict the observations.

Thereby it is imperative that an improved mathematical approach be employed to model the phenomena and deduce circulation velocity values as function of curvature. A major component which is expected to modulate the flow within the drop would be the presence of the curvature which the droplet engulfs partially. This would establish itself as a potential bluff body which modulates the flow behavior within the drop, which is otherwise similar to a plane circulation flow. The presence of the hump or depression at the droplet base distorts the otherwise plane circulation to one where the streamlines typically correspond to flow over wedges. Consequently, one approach could be to model the flow behavior via potential flow theory over angular deformations (commonly called wedge flows, with the curvature being approximated as a wedge). Fig. S4[a] illustrates the method of approximation of a curved interface to an equivalent wedge of included angle α (in radians). For wedge flows, the complex flow potential ω can be expressed as a function of z ($z=re^{i\theta}$), where $r$ and $\theta$ represent the radial and angular components of a polar coordinate system with origin at the center of the curved body, Fig. S4 [a]) coordinates in an argand plane as

$$\omega(\varphi, \psi) = Vz(r, \theta)^m \quad (2)$$

Where the index *m* represents the nature of the wedge flow as m= π / α and V is a real constant.

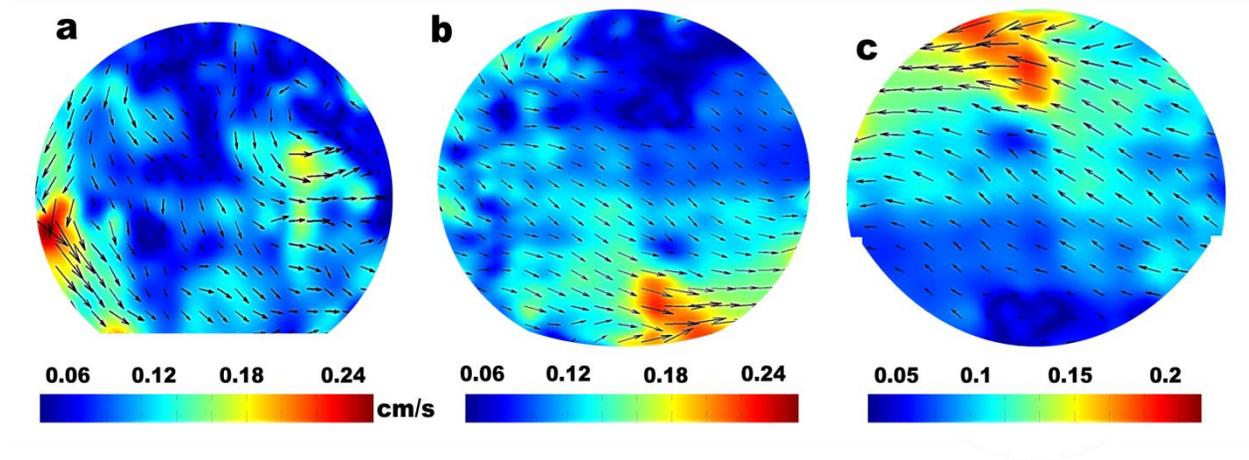

**Figure 3:** Velocity contours and vector plots for internal circulation in droplets on (a) flat (b) 3 mm concave cylinder (c) 1 mm concave cylinder

The complex flow potential can be further decomposed into the velocity potential (φ) and stream function (ψ) as $\omega = \varphi + i\psi$. From definition, the radial ($U_{m,r}$) and angular ($U_{m,\theta}$) components of circulation velocity in the droplet seated on a wedge-approximated-curvature of angle *m* can be expressed as

$$U_{m,r}(r, \theta) = mVr^{m-1}\cos(m\theta) \quad (3)$$

$$U_{m,\theta}(r,\theta) = -mVr^{m-1}\sin(m\theta) \tag{4}$$

To introduce the effect of curvature (with respect to the center of the curved surface), the variable $r$ is non-dimensionalised ($r^*=r/d$) with the distance between the centers of the droplet and the curved surface ($d=a+r_d$ for convex and $d=a-r_d$ for concave, where $r_d$ is the droplet radius). Upon introducing $r^*$, Eqns. 3 and 4 possess the dimension of velocity in the left hand, thereby forcing that the variable $V$ is also a similar term. It is scaled here that the variable is the spatially averaged circulation velocity for a flat SH case. In order to evaluate for the spatially averaged components of velocity, the expressions in Eqns. 5 and 6 are integrated with respect to radial direction (from the droplet-curvature interface to the apex of the droplet) and with respect to angular direction (from the left hand engulfing angle, $\beta_1$, to the right hand engulfing angle, $\beta_2$) (fig. S4 [a]). The present system fails for a droplet radius equal to concave curvature radius and this is justified, as the walls of the groove touches the droplet, rendering the notion of hydrophobicity geometrically void.

$$\frac{\overline{U_{m,r}}}{U_f} = \frac{m}{2r_d(\beta_2-\beta_1)} \left( \int_a^{a+2r_d} r^{*m-1}\,dr^* \right) \left( \int_{\beta_1}^{\beta_2} \cos(m\theta)d\theta \right) \tag{5}$$

$$\frac{\overline{U_{m,\theta}}}{U_f} = \frac{-m}{2r_d(\beta_2-\beta_1)} \left( \int_a^{a+2r_d} r^{*m-1}\,dr^* \right) \left( \int_{\beta_1}^{\beta_2} \sin(m\theta)d\theta \right) \tag{6}$$

Upon evaluating Eqns. 5 and 6, the expressions for a convex system can be expressed as

$$\left|\frac{\overline{U_{m,r}}}{U_f}\right| = \frac{(\theta'/\theta)^n}{2mr_d(\beta_2-\beta_1)} \left[\frac{(a+2r_d)^m - a^m}{(a+r_d)^{m-1}}\right] [sin(m\beta_2) - sin(m\beta_1)] \tag{7}$$

$$\left|\frac{\overline{U_{m,\theta}}}{U_f}\right| = \frac{(\theta'/\theta)^n}{2mr_d(\beta_2-\beta_1)} \left[\frac{(a+2r_d)^m - a^m}{(a+r_d)^{m-1}}\right] [cos(m\beta_2) - cos(m\beta_1)] \tag{8}$$

For a concave system, the limits of integration for the radial component will be $-a$ to $(2r_d - a)$. To incorporate the effects of change in wettability, the scaled contact angle ratio is introduced into the equations. The mean radial and angular components of velocity are determined from Eqn. 5 and 6. The spatially average velocity is determined as $U_m = \sqrt{U_{m,r}^2 + U_{m,\theta}^2}$. It is observed that for single curvature surfaces, $n=1$, while for dual curvatures; $n=2$. Thus, essentially, $n$ represents the number of curvatures. The deduced $U_m$ are compared with respect to experimental values in Fig. 4 and good predictability is observed for the model.

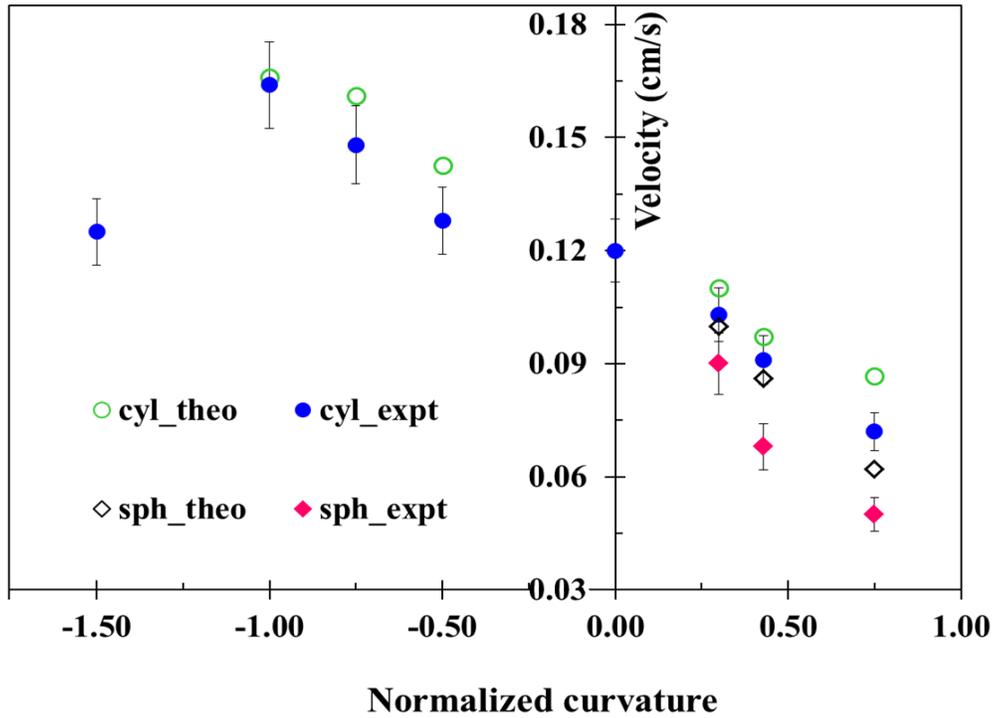

**Figure 4:** Experimental and theoretical values of the average circulation velocities. The normalized curvature is the ratio of droplet radius to curved surface radius.

Fig. S5[a] illustrates the vorticity contours within the droplets for different curvatures. It is observed that the flat SH surface (Fig S3 (a) [a]) exhibits vortex pattern distributed uniformly, indicating stable planar circulation. For convex cylinders, the vortical zone is pushed slightly away from the surface, indicating loss of circulation span (Fig S3 (b) [a]). As curvature increases, the vortical zone is pushed away towards the droplet apex, with decay in the vortex core strength (Fig S3 (c) [a]), indicating decrease in circulation span as well as strength. For concave grooves, the increase in curvature causes the vortical zone to distribute more uniformly and intensifies the core strength, since the concave shape (Fig. S3 (d) [a]) promotes stable circulation pattern compared to the flat case. When the droplet diameter is equal to the groove diameter (Fig S3 (e) [a]), the velocity reduction due to wall friction diminishes the span and magnitude of the vortical zone.

*Conclusions* – To infer, the article discusses the phenomenon of curvature dependent behavior of internal circulation within sessile droplets on SH surfaces. PIV experiments show that the circulation velocity augments for concave SH surfaces and decreases on convex surfaces. Surfaces with dual curvature are more potent in modulating the flow dynamics than single curvatures. The velocity is observed to be function of the curvature and alteration in effective wettability on curved surfaces is proposed as a mechanism. The vorticity is also mapped from PIV data and the vertical dynamics are modified by the curvatures. The curvature is approximated as wedge and potential flow theory is employed to determine the spatially averaged velocities. The mathematical predictions are observed to be consistent with experimental observations. The present findings may have important implications in microscale thermofluidic transport phenomena.

*Acknowledgements* – The authors thank the staff of the Central Workshop, IIT Ropar, for their technical assistance in fabrication of the test sections. PD also thanks IIT Ropar for funding the present research (ISIRD grant IITRPR/Research/193 and Interdisciplinary grant IITRPR/Interdisp/CDT).**References**

a) Supporting information document

1. Abramzon, B., & Sirignano, W. A. (1989). *International journal of heat and mass transfer*, *32*(9), 1605-1618.
2. Sirignano, W. A. (1983). *Progress in Energy and Combustion Science*, *9*(4), 291-322.
3. Anna, S. L., Bontoux, N., & Stone, H. A. (2003). *Applied physics letters*, *82*(3), 364-366.
4. Thorsen, T., Roberts, R. W., Arnold, F. H., & Quake, S. R. (2001). *Physical review letters*, *86*(18), 4163.
5. Ogihara, H., Xie, J., Okagaki, J., & Saji, T. (2012). *Langmuir*, *28*(10), 4605-4608.
6. Hess, J., Bo, H., Weber, R., Ortega, I., Barraud, C., van der Schoot, B., ... & De Heij, B. (2001). *U.S. Patent No. 6,196,219*. Washington, DC: U.S. Patent and Trademark Office.
7. Minemawari, H., Yamada, T., Matsui, H., Tsutsumi, J. Y., Haas, S., Chiba, R., ... & Hasegawa, T. (2011). *Nature*, *475*(7356), 364.
8. Sirringhaus, H., Kawase, T., Friend, R. H., Shimoda, T., Inbasekaran, M., Wu, W., & Woo, E. P. (2000). *Science*, *290*(5499), 2123-2126.
9. Reichard, D. L., Cooper, J. A., Bukovac, M. J., & Fox, R. D. (1998). *Pest Management Science*, *53*(4), 291-299.
10. Murakami, S. (2006). *Journal of Hydrology*, *319*(1-4), 72-82.
11. Kim, J. (2007). *International Journal of Heat and Fluid Flow*, *28*(4), 753-767.
12. Harikrishnan, A. R., Dhar, P., Gedupudi, S., & Das, S. K. (2017). *Langmuir*, *33*(43), 12180-12192.13

13. Deng, T., Varanasi, K. K., Hsu, M., Bhate, N., Keimel, C., Stein, J., & Blohm, M. (2009). *Applied Physics Letters*, *94*(13), 133109.
14. Reyssat, M., Yeomans, J. M., & Quéré, D. (2007). *EPL (Europhysics Letters)*, *81*(2), 26006.
15. Wachters, L. H. J., & Westerling, N. A. J. (1966). *Chemical Engineering Science*, *21*(11), 1047-1056.
16. Picknett, R. G., & Bexon, R. (1977). *Journal of Colloid and Interface Science*, *61*(2), 336-350.
17. Tsai, P., Pacheco, S., Pirat, C., Lefferts, L., & Lohse, D. (2009). *Langmuir*, *25*(20), 12293-12298.
18. Ma, M., & Hill, R. M. (2006). *Current opinion in colloid & interface science*, *11*(4), 193-202.
19. Koch, K., & Barthlott, W. (2009). *Philosophical Transactions of the Royal Society of London A: Mathematical, Physical and Engineering Sciences*, *367*(1893), 1487-1509.
20. Ristenpart, W. D., Kim, P. G., Domingues, C., Wan, J., & Stone, H. A. (2007). *Physical Review Letters*, *99*(23), 234502.
21. Harris, S. R., Lempert, W. R., Hersh, L., Burcham, C. L., Saville, D. A., Miles, R. B., ... & Haughland, R. P. (1996). *AIAA journal*, *34*(3), 449-454.
22. Clift, R., Grace, J. R., & Weber, M. E. (2005). *Bubbles, drops, and particles*. Courier Corporation.
23. Mandal, D. K., & Bakshi, S. (2012). *International Journal of Multiphase Flow*, *42*, 42-51.
24. Jaiswal, V., Harikrishnan, A. R., Khurana, G., & Dhar, P. (2018). *Physics of Fluids*, *30*(1), 012113.
25. Extrand, C. W., & Moon, S. I. (2008). *Langmuir*, *24*(17), 9470-9473.
26. Viswanadam, G., & Chase, G. G. (2012). *Journal of colloid and interface science*, *367*(1), 472-477.